\documentclass[aps,showpacs,preprintnumbers,amsmath, amssymb]{revtex4}

\usepackage{float}
\usepackage{graphics,epsfig}
\usepackage{graphicx}
\usepackage{dcolumn}
\usepackage{bm}

\begin{document}
\baselineskip=0.8 cm

\title{{\bf Spontaneous scalarization of Gauss-Bonnet black holes surrounded by massive scalar fields }}
\author{Yan Peng$^{1}$\footnote{yanpengphy@163.com}}
\affiliation{\\$^{1}$ School of Mathematical Sciences, Qufu Normal University, Qufu, Shandong 273165, China}

\vspace*{0.2cm}
\begin{abstract}
\baselineskip=0.6 cm
\begin{center}
{\bf Abstract}
\end{center}

For massless scalar fields, a relation $\Delta_{n}=\frac{\sqrt{3}}{2}\pi$ for $n\rightarrow \infty$
was observed in the scalar-Gauss-Bonnet theory. In the present paper,
we extend the discussion by including a nonzero scalar field mass.
For massive scalar fields, we show that the relation $\Delta_{n}=\frac{\sqrt{3}}{2}\pi$
for $n\rightarrow \infty$ still holds.
We demonstrate this relation with both analytical
and numerical methods. The analytical analysis implies that this relation may be a
very universal behavior.

\end{abstract}

\maketitle
\newpage
\vspace*{0.2cm}

\section{Introduction}

The recent gravitational wave observation confirmed
the existence of astrophysical black holes in nature \cite{BP1,BP2,BP3}.
Future more precise detection of gravitational waves may also can be used to
examine various black holes theories.
A well known property of black holes is the no hair theorem,
which suggests that asymptotically flat black holes cannot
support the formation of scalar field hairs,
see references \cite{Bekenstein}-\cite{sn3} and
reviews \cite{Bekenstein-1,CAR}.
This no hair theorem is a very general property and it
still holds even including the non-minimally coupling
between scalar fields and the Ricci curvature \cite{nm1,nm2,nm3}.

Interestingly, it was recently found that the no hair theorem is
violated when considering a non-minimally coupling
between scalar fields and the Gauss-Bonnet invariant \cite{SGB1,SGB2,SGB3,SGB4}.
In particular, this scalar-Gauss-Bonnet coupling also provides a mechanism that a thermodynamically unstable pure
black hole evolves into a thermodynamically more stable hairy black hole \cite{SGB2,SGB3}.
This intriguing mechanism is usually called spontaneous scalarization,
which was firstly found in the background of neutron stars \cite{SGBa}.
As a further step, spontaneous scalarization of Kerr black holes was also
observed in the scalar-Gauss-Bonnet theory \cite{SGB5}.
In fact, spontaneous scalarization is a very general property and it also can
be triggered by non-minimal couplings between scalar fields and Maxwell fields
\cite{charge1,charge2}. At present, spontaneous scalarization phenomenons were observed in various completed gravity models
\cite{SGB8,SGB9,SGB10,SGB11,SGB12,SGB13,SGB14,SGB15,SGB16,charge3,charge4,charge5,charge6}.
In addition, analytical properties of black hole
spontaneous scalarization were explored in \cite{hod1,hod2,hod3}.
In particular, for massless scalar fields, an interesting relation
$\Delta_{n}=\sqrt{\bar{\eta}_{n+1}}-\sqrt{\bar{\eta}_{n}}=\frac{\sqrt{3}}{2}\pi$
was obtained through WKB approach in \cite{hod1} and this relation
is also precisely supported by numerical data in \cite{SGB3}.
Whether this relation is a universal behavior is still a question to be answered.
In this work, we plan to examine whether this relation holds for scalar fields
with nonzero scalar filed mass.

The rest of this paper is as follows. We start by introducing
a model with a massive scalar field coupled to
the Gauss-Bonnet invariant in the black hole spacetime.
For massive scalar fields, we obtain a relation,
which was firstly found in the case of massless scalar fields.
We demonstrate this relation with both
analytical and numerical methods.
The conclusions are given in the last section.

\section{Investigations on the coupling parameter through the WKB method}

We study the system with a real scalar field non-minimally
coupled to the Einstein gravity. The model is described by the Lagrange density
\cite{SGB1,SGB2,SGB3,SGB4}
\begin{eqnarray}\label{lagrange-1}
\mathcal{L}=R-\nabla^{\nu}\nabla_{\nu}\Psi-\mu^{2}\Psi^{2}+f(\Psi)\mathcal{R}_{GB}^{2}.
\end{eqnarray}
Here R is the Ricci scalar of the metric.
$\Psi$ is the scalar field with mass $\mu$.
The term $\mathcal{R}_{GB}^{2}$ is the Gauss-Bonnet invariant expressed as
$\mathcal{R}_{GB}^{2}=R_{\alpha\beta\rho\sigma}R^{\alpha\beta\rho\sigma}-4R_{\alpha\beta}R^{\alpha\beta}+R^2$.
In the Schwarzschild black hole background, it is given by $\mathcal{R}_{GB}^{2}=\frac{48M^2}{r^6}$ \cite{hod1,hod2,Yan1,Yan2}.
The function $f(\Psi)$ describes the coupling between the scalar field
and the Gauss-Bonnet invariant. In the linearized regime, without loss of generality,
the coupling function can be taken in the form $f(\Psi)=\eta\Psi^2$,
with $\eta$ as the coupling parameter \cite{hod1,hod2,Yan1,Yan2}.

The scalar field equation of motions is
\begin{eqnarray}\label{lagrange-1}
\nabla^{\nu}\nabla_{\nu}\Psi-\mu^2\Psi+\frac{f'_{\Psi}\mathcal{R}_{GB}^{2}}{2}=0.
\end{eqnarray}

The line element describing a spherically symmetric black hole solution is
\begin{eqnarray}\label{AdSBH}
ds^{2}&=&-g(r)dt^{2}+\frac{dr^{2}}{g(r)}+r^{2}(d\theta^{2}+sin^{2}\theta d\phi^{2}).
\end{eqnarray}
The metric function of a scalar free Schwarzschild black hole
is given by $g(r)=1-\frac{2M}{r}$ with M as the black hole mass.
The black hole horizon $r_{h}=2M$ is determined by the relation $g(r_{h})=0$.

According to relations (2) and (3), the scalar field equation can be expressed in the form \cite{Yan1,Yan2}
\begin{eqnarray}\label{BHg}
\Psi''+(\frac{2}{r}+\frac{g'}{g})\Psi'+(\frac{\eta\mathcal{R}_{GB}^{2}}{g}-\frac{\mu^2}{g})\Psi=0
\end{eqnarray}
with $g=1-\frac{2M}{r}$ and $\mathcal{R}_{GB}^{2}=\frac{48M^2}{r^6}$.

At the horizon $r_{h}=2M$, we take finite boundary conditions \cite{hod1}
\begin{eqnarray}\label{InfBH}
\Psi(r_{h})<\infty.
\end{eqnarray}

In the far region, the general asymptotical behavior of massive static scalar fields
are $\Psi(r\rightarrow \infty)\sim \frac{A_{1}}{r}e^{-\mu r}+\frac{B_{1}}{r}e^{\mu r}$
with $A_{1}$ and $B_{1}$ as constants.
For bounded scalar fields, we impose the relation $B_{1}=0$.
So the infinity boundary condition is \cite{hod2,Yan1,hod4,hod5}
\begin{eqnarray}\label{InfBH}
\Psi(r\rightarrow \infty)\varpropto \frac{1}{r}e^{-\mu r}.
\end{eqnarray}

We introduce a new function $\psi=r\Psi$. Then there is $\psi(r\rightarrow \infty)\rightarrow 0$.
We also take the tortoise radial coordinate y defined by the relation
\begin{eqnarray}\label{BHg}
\frac{dr}{dy}=g(r).
\end{eqnarray}

With relations (4), (7) and $\mathcal{R}_{GB}^{2}=\frac{48M^2}{r^6}$,
we derive the Schr\"{o}dinger-like differential equation
\begin{eqnarray}\label{BHg}
\frac{d^{2}\psi}{dy^{2}}-V\psi=0
\end{eqnarray}
with the effective potential expressed as

\begin{eqnarray}\label{BHg}
V(r)=(1-\frac{2M}{r})(\frac{2M}{r^3}+\mu^2-\frac{12\eta M^2}{r^6}).
\end{eqnarray}

The Schr\"{o}dinger-like radial differential
equation (8) can be studied through a standard WKB analysis.
For a bounded scalar field, there is a well-known quantization condition \cite{hod1}
\begin{eqnarray}\label{BHg}
\int_{y_{-}}^{y_{+}}dy\sqrt{-V(y,\eta)}=(n-\frac{1}{4})\pi;~~~n=1,2,3,\ldots,
\end{eqnarray}
where $y_{\pm}$ are the classical turning points derived from
$V(y_{\pm})=0$. And n is a positive integer.

With the relation (7), we can transfer the resonance
condition (10) into
\begin{eqnarray}\label{BHg}
\int_{r_{-}}^{r_{+}}dr\frac{\sqrt{-V(r,\eta)}}{g(r)}=(n-\frac{1}{4})\pi;~~~n=1,2,3,\ldots,
\end{eqnarray}
where the two turning points $r_{\pm}$ are determined by $V(r_{\pm})=0$ from
equations
\begin{eqnarray}\label{BHg}
1-\frac{2M}{r_{-}}=0
\end{eqnarray}
and
\begin{eqnarray}\label{BHg}
\frac{2M}{r_{+}^3}+\mu^2-\frac{12\eta M^2}{r_{+}^6}=0.
\end{eqnarray}

Then the turning points are
\begin{eqnarray}\label{BHg}
r_{-}=2M,~~~~~~r_{+}=(\frac{12\eta}{1+ \sqrt{1+12\eta \mu^2}})^{1/3}.
\end{eqnarray}

Substituting (9) into equation (11), we obtain the relation
\begin{eqnarray}\label{BHg}
\int_{r_{-}}^{r_{+}}dr\sqrt{\frac{12\eta M^2-\mu^2r^6-2Mr^3}{r^6(1-\frac{2M}{r})}}=(n-\frac{1}{4})\pi;~~~n=1,2,3,\ldots.
\end{eqnarray}
For fixed values of $\eta$, $\mu$ and $M$, the integration on the left side of (15) is finite.
From the relation (15), for fixed $\mu$ and $M$, there is $\eta\rightarrow \infty$ in the limit of $n\rightarrow \infty$.
Also considering (14), it yields $r_{+}\rightarrow \infty$ in the limit of $n\rightarrow \infty$.
In the large n limit, the left side of (15) can be approximated by
\begin{eqnarray}\label{BHg}
\int_{r_{-}}^{r_{+}}dr\sqrt{\frac{12\eta M^2-\mu^2r^6-2Mr^3}{r^6(1-\frac{2M}{r})}}=\int_{2M}^{\infty}dr\sqrt{\frac{12\eta M^2}{r^6(1-\frac{2M}{r})}}.
\end{eqnarray}

According to (16), the relation (15) can be expressed as
\begin{eqnarray}\label{BHg}
\int_{2M}^{\infty}dr\sqrt{\frac{12\bar{\eta} M^4}{r^6(1-\frac{2M}{r})}}=(n-\frac{1}{4})\pi~~~for~~~n\rightarrow \infty,
\end{eqnarray}
where we define a new dimensionless parameter $\bar{\eta}=\frac{\eta}{M^2}$
using the symmetry $r\rightarrow \gamma r,~ \mu\rightarrow \mu/\gamma,~ M\rightarrow \gamma M,~ \eta\rightarrow \gamma^2 \eta$ of equation (4).
According to the integration $\int_{2M}^{\infty}dr\sqrt{\frac{12 M^4}{r^6(1-\frac{2M}{r})}}=\frac{2}{\sqrt{3}}$,
one deduces the expression
\begin{eqnarray}\label{BHg}
\sqrt{\bar{\eta}_{n}}=\frac{\sqrt{3}}{2}(n-\frac{1}{4})\pi.
\end{eqnarray}

The discrete resonant spectrum (18) yields a relation
\begin{eqnarray}\label{BHg}
\Delta_{n}=\sqrt{\bar{\eta}_{n+1}}-\sqrt{\bar{\eta}_{n}}=\frac{\sqrt{3}}{2}\pi.
\end{eqnarray}
We mention that the relation (19) is the same as the case of massless
scalar fields in \cite{hod1}. Our analysis implies that this relation may hold
for very general models with the property that
$r_{+}\rightarrow \infty$ as $n\rightarrow \infty$.
Since the WKB approach is an approximate method,
we still have to examine the relation with numerical data.

\section{ Numerical examination of the universal behavior}

In this part, we examine the relation (19) with numerical solutions
by integrating the equation (4) from the black hole horizon to the infinity.
Using the symmetry $\psi\rightarrow \kappa \psi$ of equation (4),
we can take $\psi(r_{h})=1$ without loss of generality.
From mathematical aspects, it is a boundary value problem.
For fixed parameters $\mu^2$ and $M$, we find
discrete $\eta$, which leads to bounded scalar field solutions
satisfying the boundary conditions $\psi(r_{h})=1$ and $\psi(\infty)=0$.

With $\mu^2=0.002$ and $M=1$, we choose different values of $\eta$
to try to search for the proper $\eta$ corresponding to a
scalar field decaying at the infinity.
We numerically obtain a discrete value $\eta\thickapprox3.110$,
which leads to the solution with $\Psi(\infty)=0$.
We plot the physical solution with blue curves in the left panel of Fig. 1.
With a larger discrete value $\eta\thickapprox20.263$,
we get a solution with one node as shown in the right panel of Fig. 1.
In fact, we can also get physical solutions with many nodes.

\begin{figure}[h]
\includegraphics[width=180pt]{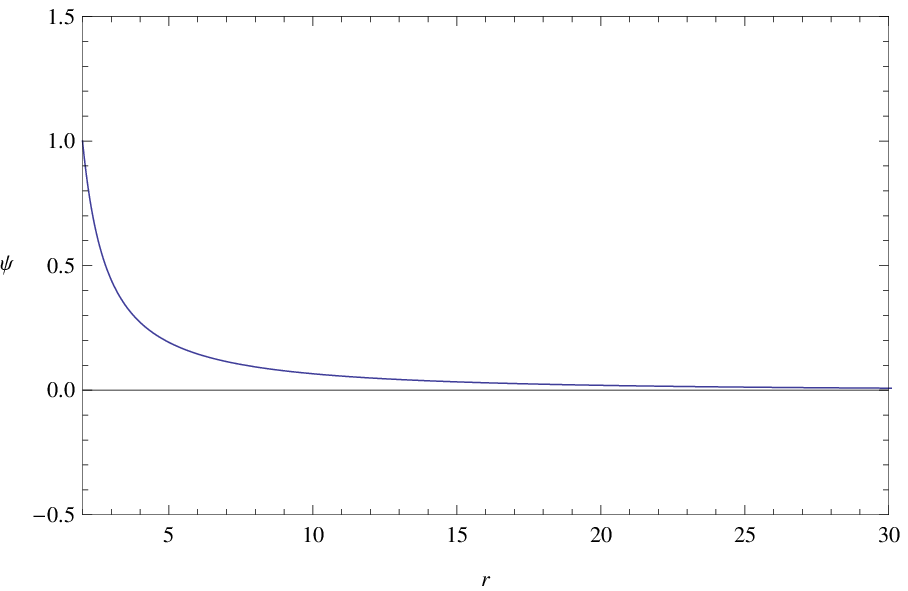}\
\includegraphics[width=180pt]{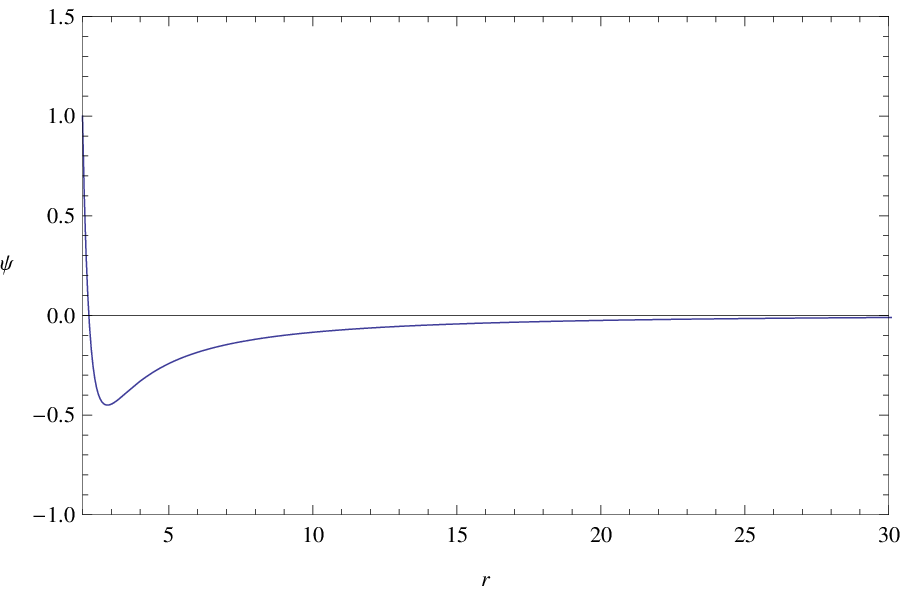}\
\caption{\label{EEntropySoliton} (Color online) We show the scalar field $\Psi(r)$
as a function of the coordinate r with $\mu^2=0.002$ and $M=1$. The left panel corresponds to the case of
$\eta=3.110$ and the right panel is with $\eta=20.263$.}
\end{figure}

With numerical solutions, we can examine whether the relation (19) holds.
We study effects of scalar field mass $\mu^2$ and node number n
on the discrete dimensionless coupling parameter $\bar{\eta}(\mu^2,n)$.
With $M=1$, we choose $\mu^2=0.002$ in Table I and $\mu^2=0.004$ in Table II.
The data shows that, for nonzero scalar field mass $\mu^2$,
$\Delta_{n}=\sqrt{\bar{\eta}_{n+1}}-\sqrt{\bar{\eta}_{n}} \rightarrow 2.72$
for very large n. Since $\frac{\sqrt{3}}{2}\pi \thickapprox 2.72\times 1.0003$, we have numerically
demonstrated the analytical relation (19).
For larger mass $\mu^2$, we usually need a
larger n to obtain the limit value $\bar{\eta}\thickapprox 2.72$,
as can be seen from Table I and Table II.

\renewcommand\arraystretch{2.0}
\begin{table} [h]
\centering
\caption{The parameter $\bar{\eta}$ with $\mu^2=0.002$ and $M=1$}
\label{address}
\begin{tabular}{|>{}c|>{}c|>{}c|>{}c|>{}c|>{}c|}
\hline
$~n~$ & ~1~& ~2~& ~3~& ~$4\leqslant n\leqslant 28$~& ~$n\geqslant 29$\\
\hline
$~\bar{\eta}~$ & ~2.74~& ~2.74~& ~2.74~& ~2.73~& ~2.72\\
\hline
\end{tabular}
\end{table}

\renewcommand\arraystretch{2.0}
\begin{table} [h]
\centering
\caption{The parameter $\bar{\eta}$ with $\mu^2=0.004$ and $M=1$}
\label{address}
\begin{tabular}{|>{}c|>{}c|>{}c|>{}c|>{}c|>{}c|}
\hline
$~n~$ & ~1~& ~2~& ~$3\leqslant n\leqslant 5$~& ~$6\leqslant n\leqslant 41$~& ~$n\geqslant 42$\\
\hline
$~\bar{\eta}~$ & ~2.75~& ~2.75~& ~2.74~& ~2.73~& ~2.72\\
\hline
\end{tabular}
\end{table}

\section{Conclusions}

We investigated formations of scalar clouds outside
asymptotically flat spherical black holes.
We considered massive scalar fields non-minimally coupled to
the Gauss-Bonnet invariant. For massive scalar fields, with WKB methods, we analytically
obtained a relation $\Delta_{n}=\sqrt{\bar{\eta}_{n+1}}-\sqrt{\bar{\eta}_{n}}=\frac{\sqrt{3}}{2}\pi$
in the limit of $n\rightarrow \infty$.
We numerically found $\Delta_{n}=\sqrt{\bar{\eta}_{n+1}}-\sqrt{\bar{\eta}_{n}}\rightarrow 2.72$
in the limit of $n\rightarrow \infty$.
Considering the fact $\frac{\sqrt{3}}{2}\pi \thickapprox 2.72\times 1.0003$,
we numerically demonstrated the relation
$\Delta_{n}=\sqrt{\bar{\eta}_{n+1}}-\sqrt{\bar{\eta}_{n}}=\frac{\sqrt{3}}{2}\pi$ in the large n limit.
In fact, this relation was firstly obtained in the case of massless scalar fields in \cite{hod1}
and supported by numerical data in \cite{SGB3}.
We also pointed out that the WKB analysis could be generalized
to more general models, which implies that the relation may be a
very universal behavior.

\begin{acknowledgments}

This work was supported by the Shandong Provincial Natural Science Foundation of China under Grant
No. ZR2018QA008. This work was also supported by a grant from Qufu Normal University of China under Grant
No. xkjjc201906.

\end{acknowledgments}

\end{document}